\documentstyle[12pt]{article}

\newcommand{\remove}[1]{}
\newtheorem{theorem}{Theorem}[section]
\newtheorem{lemma}[theorem]{Lemma}
\newtheorem{claim}[theorem]{Claim}

\newcommand{\proof}{ \noindent {\bf Proof: \ } \\ }
\def\QED {\hfill $\!\Box$ \newline}

\def\Agr{{\mathrm Agr}}
\def\QPCP{{\mathrm QPCP}}
\def\P{{\mathrm Prob}}
\def\E{{\mathrm Exp}}

\def\H{{\mathrm H}}

\date{}

\begin{document}

\title{Quantum Information and the PCP Theorem}

\author{
Ran Raz\thanks{Research supported by Israel Science Foundation
(ISF) grant. }\\
Weizmann Institute \\
{\tt ran.raz@weizmann.ac.il}
}

\maketitle

\thispagestyle{empty}

\begin{abstract}
We show how to encode $2^n$ (classical) bits
$a_1,...,a_{2^n}$ by a single quantum state $|\Psi \rangle$
of size $O(n)$ qubits, such that:
for any constant $k$ and any
$i_1,...,i_k \in \{1,...,2^n\}$,
the
values of the bits $a_{i_1},...,a_{i_k}$ can be retrieved
from $|\Psi \rangle$ by a one-round
Arthur-Merlin interactive protocol of size polynomial in $n$.
This shows how to go around Holevo-Nayak's Theorem, using
Arthur-Merlin proofs.

We use the new representation to prove the following results:
\begin{enumerate}
\item
Interactive proofs with quantum advice:\\
We show that the class $QIP/qpoly$ contains {\bf all} languages.
That is,
for any language $L$ (even non-recursive),
the membership $x \in L$ (for $x$ of length $n$)
can be proved by a polynomial-size
quantum interactive proof,
where
the verifier is a polynomial-size quantum circuit
with working space initiated with some
quantum state $|\Psi_{L,n} \rangle$
(depending only on $L$ and $n$).
Moreover, the interactive proof that we give
is of only one round,
and the messages communicated are classical.
%(i.e.,
%a polynomial-size quantum state $|\Psi \rangle$,
%depending only on $n$,
%is given to the verifier as an advice.
\item
PCP with only one query:\\
We show that
the membership $x \in SAT$ (for $x$ of length $n$)
can be proved by a logarithmic-size
quantum state $|\Psi \rangle$, together with a
polynomial-size classical proof consisting of
blocks of length $polylog(n)$ bits each,
such that after measuring the state $|\Psi \rangle$
the verifier only needs to read {\bf one} block
of the classical proof.
\end{enumerate}
While the first result is a straight forward consequence of the
new representation, the second requires an additional machinery of
{\em quantum low-degree-test} that may be interesting
in its own right.
\end{abstract}

\section{Introduction}

\subsection{Around Holevo's Theorem} \label{subsec:Holevo}

A quantum state of $n$ qubits contains an infinite amount
of information. If the state is only given up to
some fixed (say, constant) precision it still contains
an exponential amount of information.
On the other hand, a quantum measurement can only give
$n$ bits of information about the state.
One way to formalize the last statement is given by Holevo's
theorem~\cite{Hol}.

A simplified version of Holevo's theorem can be stated
as follows:
If $n$ (classical) bits $a_1,...,a_n$ are encoded
by a single quantum state
$|\Psi \rangle =|\Psi(a_1,...,a_n) \rangle$,
such that
the original values of the
bits $a_1,...,a_n$ can be retrieved from the state
$|\Psi \rangle$,
then $|\Psi \rangle$ is a state of
at least $n$ qubits.
In other words:
Assume that Bob encodes $n$ bits
$a_1,...,a_n$ by a quantum state
$|\Psi \rangle$ and sends
$|\Psi \rangle$ to Alice.
Assume that Alice can retrieve the original values of
$a_1,...,a_n$ by measuring the state
$|\Psi \rangle$. Then,
$|\Psi \rangle$ is a state of
at least $n$ qubits.
Moreover, if we only require that each $a_i$ is retrieved
correctly with probability $1- \epsilon$, and allow
an error to occur with probability $\epsilon$,
then $|\Psi \rangle$ is a state of
at least $(1-\H(\epsilon)) \cdot n$ qubits,
where $\H(\epsilon)$ denotes the Shannon's entropy of the
distribution $(\epsilon,1-\epsilon)$.

A strengthening of Holevo's theorem was suggested
by Ambainis, Nayak, Ta-Shma, and Vazirani~\cite{ANTV}
and was proved by Nayak~\cite{Nay}.
A simplified version of Nayak's theorem can be stated
as follows:
Assume that Bob encodes $n$ bits
$a_1,...,a_n$ by a quantum state
$|\Psi \rangle$ and sends
$|\Psi \rangle$ to Alice.
Assume that for every index $i \in \{1,...,n\}$
(of her choice),
Alice can retrieve the original value of
$a_i$ by measuring the state
$|\Psi \rangle$. Then,
$|\Psi \rangle$ is a state of
at least $n$ qubits.
Moreover, if we only require that Alice retrieves $a_i$
correctly with probability $1- \epsilon$, and allow
an error to occur with probability $\epsilon$,
then $|\Psi \rangle$ is a state of
at least $(1-\H(\epsilon)) \cdot n$ qubits.

Note that the difference between Holevo's theorem and Nayak's
theorem is that in Holevo's theorem we require that Alice can
retrieve the values of {\bf all} the original bits,
whereas in Nayak's theorem we only require
that Alice can retrieve the value of {\bf one} bit of her choice.
Note that by the uncertainty principle these two tasks are
not necessarily equivalent. It was demonstrated
in~\cite{ANTV} that the two tasks are indeed not equivalent.

In this paper, we suggest the use of Arthur-Merlin protocols
to go around Holevo's and Nayak's theorems.
Roughly speaking:
Bob will encode a large number of (classical) bits by a
very short quantum state and will send that state to Alice.
Alice will not be able to retrieve even one of the original
bits by herself. Nevertheless, the value of each one of the
original bits can be retrieved by an Arthur-Merlin protocol,
with a third party, the infinitely powerful prover Merlin.
In this protocol, Alice acts as the verifier Arthur.
In other words, although Alice is not able to retrieve
the value of the $i$th bit by herself, Merlin will
tell her that value and will be able to convince her that
this value is correct.
Note that in this setting Bob is completely trustable and hence
Alice can count that the quantum state given by Bob
correctly encodes
the original bits. Merlin, on the other hand, cannot be trusted
and hence Alice needs to be convinced that his answer is correct.

Interestingly, the communication between Alice and Merlin
in our protocol
will be {\bf classical}. They will not need to exchange quantum
states. We can hence assume w.l.o.g.
that Merlin is an infinitely powerful {\bf classical} computer.
Alice, on the other hand, will need to have
the ability to measure the quantum state sent by Bob,
but her computational power will be polynomially bounded
(as required in an Arthur-Merlin protocol).

More precisely,
we will construct a protocol that works as follows:
Bob encodes $2^n$ (classical) bits
$a_1,...,a_{2^n}$ by a quantum state
$|\Psi \rangle = |\Psi(a_1,...,a_{2^n}) \rangle$
of size $O(n)$ qubits,
and sends $|\Psi \rangle$
to Alice.
Alice measures the state $|\Psi \rangle$.
Given an index $i \in \{1,...,2^n\}$
(of her choice), and based on the result of the measurement,
Alice composes a (classical) question $q$ of length
$poly(n)$ bits and sends $(i,q)$ to Merlin.
After seeing $(i,q)$, Merlin responds with a (classical)
answer $r$ of length $poly(n)$ bits.
Based on $i,q,r$, Alice decides on a value
$V(i,q,r) \in \{0,1,Err\}$, where $0$ is interpreted
as $a_i=0$ and $1$ is interpreted
as $a_i=1$, and $Err$ is interpreted as a declaration
that Merlin is cheating.
We will have the following (standard) completeness and soundness
properties for this protocol:
\begin{enumerate}
\item
For any $i,q$, there is an answer $r$, such that
$V(i,q,r) = a_i$ (with probability 1).
\item
For any $i,q,r$, we have that $V(i,q,r) \in \{a_i,Err\}$ with
probability $\geq 1- 1/n^{\Omega(1)}$.
\end{enumerate}
In other words, for any index $i$ and question $q$, Merlin
will be able to give an answer $r$ that causes Alice
to conclude the correct value of $a_i$, and on the other
hand, no answer given by Merlin can cause Alice to
conclude the incorrect value of $a_i$
(with non-negligible probability).

Our results are in fact more general:
We will be able to encode and retrieve
$a_1,...,a_{2^n}$ that can get
$n^{O(1)}$ different values, rather than bits
(i.e., each $a_i$ can be a block of $O(\log n)$ bits).
Moreover, we will be able to retrieve any constant
number of values $a_{i_1},...,a_{i_k}$.

\subsection{The Exceptional Power of $QIP/qpoly$} \label{subsec:QAM}

Interactive proofs were introduced by Goldwasser, Micali
and Rackoff and by Babai and Moran~\cite{GMR,Bab,BM},
and were extended to the quantum case by Watrous~\cite{Wat}.
The simplest version of a quantum interactive proof is
a {\bf one-round} (i.e., two messages)
quantum interactive proof, usually called a $QIP(2)$ proof.

In a $QIP(2)$ proof,
the infinitely powerful prover Merlin
tries to convince the verifier
Arthur for a membership $x \in L$,
(where $L$ is some language and $x$ is an input
of length $n$, and
both $x$ and $L$ are known to both parties).
Both parties have quantum computers
and they can communicate between them quantum states.
Merlin's computational power is unlimited
(but he must obey the laws of physics).
Arthur's computational power, on the other hand,
is limited to (quantum) polynomial time.
The proof has one round of communication,
where the two parties exchange between them quantum states.

In this paper, we will not need the full power
of $QIP(2)$ proofs. We will use a subclass of proofs
that we call $QIP(2)^*$ proofs.
In a $QIP(2)^*$ proof,
Arthur and Merlin communicate between
them {\bf classical} messages, rather than quantum states.
We can hence assume w.l.o.g. that Merlin is an infinitely
powerful {\bf classical} computer. Arthur, on the other hand,
will need to have the ability to work with quantum states
(in order to be able to work with the quantum advice
discussed below).

A $QIP(2)^*$ proof has one round of communication:
Based on $x$ (and possibly on a random string),
Arthur composes a classical question $q$ of length
$poly(n)$ bits and sends $q$ to Merlin.
After seeing $(x,q)$, Merlin responds with a classical
answer $r$ of length $poly(n)$ bits.
Based on $x,q,r$, Arthur decides on a value
$V(x,q,r) \in \{Accept,Reject\}$, where $Accept$ is interpreted
as $x \in L$ and $Reject$ is interpreted
as a declaration that Merlin is cheating.
The following completeness and soundness
properties should be satisfied\footnote{In the
protocols constructed in this paper, we will actually have
perfect completeness, i.e., $\epsilon =0$ in the first
item bellow.} (for some small
constant $\epsilon$):
\begin{enumerate}
\item
For any $x \in L$ and any $q$, there is an answer $r$, such that
$V(x,q,r) = Accept$, with probability $\geq 1-\epsilon$.
\item
For any $x \not \in L$ and any $q,r$,
we have that $V(x,q,r) = Reject$, with
probability $\geq 1- \epsilon$.
\end{enumerate}
In other words, if $x \in L$ then for any question $q$ Merlin
will be able to give an answer $r$ that causes Arthur
to accept (with high probability), and on the other
hand,
if $x \not \in L$ then for any question $q$
no answer given by Merlin can cause Arthur to
accept (with non-negligible probability).

In this paper, we are interested in the class $QIP(2)^*/qpoly$,
that is, the class of languages that have polynomial-size
$QIP(2)^*$ proofs with a {\bf polynomial-size quantum advice}.
A $QIP(2)^*/qpoly$ proof is the same as a $QIP(2)^*$ proof,
except that the computational power of Arthur
is quantum polynomial time with a polynomial-size quantum advice.
In other words, Arthur is a quantum circuit in
$BQP/qpoly$.
We can think of a circuit in $BQP/qpoly$ as a
polynomial-size quantum circuit
with working space initiated with an arbitrary
quantum state $|\Psi_{L,n} \rangle$
(depending only on $L$ and~$n$).
We think of the state $|\Psi_{L,n} \rangle$
as a (polynomial-size) {\em quantum advice}
(given to the verifier).

The notion of quantum advice was studied in several previous
works~\cite{NY,Aar}, as a quantum analog to the notion
of classical advice (or classical non-uniformity).
These works concentrated on the class $BQP/qpoly$
and proved some limitations of that class.
In particular,
Aaronson proved that the class $BQP/qpoly$ is
contained in the classical class $PP/poly$~\cite{Aar}.
It is hence somewhat surprising that
$QIP(2)^*/qpoly$ proofs are so powerful.

We show that the class $QIP(2)^*/qpoly$ contains {\bf all}
languages.
That is,
for any language $L$, there is a polynomial-size
$QIP(2)^*/qpoly$ interactive
proof for the membership $x \in L$.
Since any $QIP(2)^*/qpoly$ proof is also
a $QIP(2)/qpoly$ proof, this obviously means that
that the class $QIP(2)/qpoly$,
and hence also $QIP/qpoly$,
contain all languages.

\subsection{A Quantum Version of the PCP Theorem} \label{subsec:PCP}

A probabilistic checkable proof (PCP) is a proof that
can be (probabilistically) verified by reading only
a small portion of it.
The PCP theorem~\cite{BFL,FGLSS,AS1,ALMSS} states that
for any $x \in SAT$ (where $x$ is an input of length $n$), there is
a PCP $p$ for the membership $x \in SAT$, such that
the proof $p$ is of length $poly(n)$ bits
and it can be (probabilistically) verified by reading
only a constant number of its bits.
Moreover, there is
a PCP $p$ for the membership $x \in SAT$, such that
the proof $p$ is consisted of $poly(n)$ blocks of length $O(1)$
bits each
and it can be (probabilistically) verified by reading
only {\bf two} of its blocks.
A similar PCP that can be verified by reading
only one of its blocks is obviously impossible, under
standard hardness assumptions (even if we allow the length
of each block to be almost linear).

In this paper, we show that
the membership $x \in SAT$ (for $x$ of length $n$)
can be proved by a logarithmic-size
quantum state $|\Psi \rangle$, together with a
polynomial-size classical proof $p$ consisting of
blocks of length $polylog(n)$ bits each,
such that after measuring the state $|\Psi \rangle$
the verifier only needs to read {\bf one} block
of the proof $p$.

More precisely, the verifier can be modelled by
a polynomial-size quantum circuit.
For any $x \in SAT$,
there exists a logarithmic-size quantum state $|\Psi \rangle$
and an array $p$ of $poly(n)$ blocks of length
$polylog(n)$ bits each, that encode a proof
for the membership $x \in SAT$ and
can be verified as follows:
The verifier applies on $|\Psi \rangle$ a
carefully designed (probabilistic) unitary transformation
$U$ (that can be computed in quantum logarithmic time).
The verifier measures some of the bits of $U|\Psi \rangle$.
Denote the collapsed state (after the measurement) by
$|\Psi' \rangle$.
Based on $x$ and on the result of the measurement,
the verifier composes a (classical) query $q$
(of length $O(\log n)$ bits)
and reads the $q$th block of $p$.
Denote the value of that block by $r$.
Based on $x,q,r$, the verifier applies a unitary transformation
$U'$ (that can be computed in quantum logarithmic time)
on $|\Psi' \rangle$ and measures all bits of
$U'|\Psi' \rangle$.
Based on the result of the measurement, the verifier
decides whether to $Accept$ or $Reject$,
where $Accept$ is interpreted
as $x \in SAT$ and $Reject$ is interpreted
as a declaration that the proof $(|\Psi \rangle,p)$ is not correct.
We will have the following completeness and soundness
properties (for any fixed constant $\epsilon > 0$):
\begin{enumerate}
\item
For any $x \in SAT$, there
exist $|\Psi \rangle$ and $p$ that cause the verifier
to accept with probability~1.
\item
For any $x \not \in SAT$, and any $|\Psi \rangle$ and $p$,
the verifier rejects with probability $\geq 1- \epsilon$.
\end{enumerate}

\subsection{Methods}

We combine methods
previously used in the field of
probabilistic checkable proofs and methods previously used
in the field of quantum computations, together with
some new ideas.

The quantum state $|\Psi(a_1,...,a_{2^n}) \rangle$, from
Subsection~\ref{subsec:Holevo}, is a quantum representation
of the so called, {\em low degree extension}, of $a_1,...,a_{2^n}$.
Low degree extensions were extensively used in the past
in the study of randomness and derandomization
and probabilistic checkable proofs.
For the retrieval protocol of Subsection~\ref{subsec:Holevo},
we use the
random self reducibility property and the
locally decodability property of the low degree extension.

For the results discussed in Subsection~\ref{subsec:QAM},
we will use as a quantum advice the quantum state
$|\Psi(a_{0},...,a_{2^{n}-1}) \rangle$, where
$a_i =1$ iff $i \in L$.
The results of Subsection~\ref{subsec:QAM} will then
follow immediately from the ones of
Subsection~\ref{subsec:Holevo}.

For the results discussed in Subsection~\ref{subsec:PCP},
we will use the quantum state
$|\Psi(a_1,...,a_{m}) \rangle$, where
$(a_1,...,a_{m})$ is a (classical) PCP for the
membership $x \in SAT$.
Note, however, that in the setting of Subsection~\ref{subsec:PCP}
the verifier cannot assume anything about the quantum
state given to him, as it is given by the (un-trusted) prover.
The verifier
cannot even trust that the quantum
state given to him is a correct representation
of the low degree extension of some sequence of bits.
A key step in our analysis will be a
{\em quantum low degree test} that will
ensure that the state is close to a quantum representation
of some multivariate polynomial of low degree.
Since this seems to be impossible
for the verifier to do by himself, the test is done with
the help of a classical PCP (or equivalently, with the help
of a classical prover).

Note that in the setting of Subsection~\ref{subsec:PCP},
the verifier cannot query the classical proof more than once.
Moreover, the verifier can measure the quantum state only once
(as the state collapses after the measurement).
Hence, the verifier cannot apply both the quantum
low degree test and the retrieval protocol.
We will hence need to integrate these two tasks.
We will do that using ideas from~\cite{DFKRS}.
A special attention is given to the probability of error,
as we would like to keep it as small as possible
(and in particular, sub-constant).

Most of the technical work in the paper is done in the proofs
of the results
discussed in Subsection~\ref{subsec:PCP}
(including the proof for the correctness of the quantum low
degree test).

\subsection{Discussion}

The PCP style results of Subsection~\ref{subsec:PCP}
scale up to languages in $NEXP$.
More precisely,
for any language $L \in NEXP$,
the membership $x \in L$ (for $x$ of length $n$)
can be proved by a polynomial-size
quantum state $|\Psi \rangle$, together with an
exponential-size classical proof $p$ consisting of
blocks of length $poly(n)$ bits each,
such that after measuring the state $|\Psi \rangle$
the verifier only needs to read one block
of the proof $p$.

There are several alternative ways to present the last result.
One of them is the following:
Consider a two-rounds interactive proofs model, where
the prover has {\bf quantum} power in the first round but
only {\bf classical} power in the second round
(note that in the second round the prover still has an infinitely
powerful classical computer, but he cannot access any quantum state).
Then, for any language $L \in NEXP$,
the membership $x \in L$ (for $x$ of length $n$)
can be proved by a polynomial-size interactive proof in this model.

Note that $IP=PSPACE$~\cite{LFKN,Sha}, and $QIP \subset EXP$~\cite{KW}.
Thus, if the prover has classical power in both rounds or quantum
power in both rounds we are not likely to be able to
prove memberships $x \in L$ even for languages
$L \in NTIME(n^{\log n})$.
In contrast, if the prover has quantum power in the first round
and classical power in the second we are able to prove memberships
$x \in L$ for any $L \in NEXP$.

One can ask why it is not possible to use the same protocol
when the prover is quantum in both rounds.
The reason is that if we do so,
the answers given by the prover in the second round
may depend on the results of a measurement of a quantum state that
is entangled to the state supplied to the verifier
in the first round. This forms a sophisticated version
of the EPR paradox, in the spirit of~\cite{CHTW}.

\subsection{Preliminaries}

We assume that the reader is familiar with the basic concepts
and notations
of quantum computation. For excellent surveys on the subject
see~\cite{Aha,NC}.

Let $F$ be a field of size $2^a$ for some integer $a$
(that will be a function of $n$ and will be determined later on).
Our basic quantum element will be a quantum register of $a$ qubits,
rather than a single qubit.
Each such basic element represents a member of the
Hilbert space $C^{|F|}$.
We denote by $\{ |e \rangle \}_{e \in F}$
the standard basis for that space.

The base for the logarithm in this paper is always 2.
By $[m]$ we denote the set $\{1,...,m\}$.
We denote probabilities by $\P$ and expectations by $\E$.
We say that a multivariate polynomial is of total degree $r$
if its total degree is {\bf at most} $r$.

\section{The Exponential Information of a Quantum State}

In this section, we present the results discussed
in Subsection~\ref{subsec:Holevo}.
We will encode $2^n$ (classical) bits
$a_1,...,a_{2^n}$ by a quantum state
$|\Psi \rangle = |\Psi(a_1,...,a_{2^n}) \rangle$
of size $O(n)$ qubits.
We will show how to retrieve the value of any of the
original bits by a (polynomial-size) Arthur-Merlin protocol.
Our protocol is in fact more general:
We will be able to encode and retrieve
$a_1,...,a_{2^n}$ that can get
$n^{O(1)}$ different values, rather than bits
(i.e., each $a_i$ can be a block of $O(\log n)$ bits).

\subsection{Quantum Low Degree Extension} \label{subsec:qlde}

W.l.o.g., assume that $n > 4$ is an even power of 2
(otherwise, we just increase $n$ to at most $4n$,
by padding with zeros).
Denote by $F$ a field of size $2^a \doteq n^c$, where $c$
is a large enough constant integer that will be determined later on
(for the content of this section, $c=2$ is enough).
Let $H \subset F$ be any (efficiently enumerable)
subset of size $\sqrt{n}$
(e.g., the lexicographically first elements in some representation
of the field $F$).
Denote $d = 2n/\log n$, and assume for simplicity
of the presentation that $d$ is integer.
Note that $|H^d| = 2^n$.
Denote by $\pi : H^d \rightarrow [2^n] $  any
(efficiently computable)
one-to-one function (e.g., the lexicographic order of $H^d$).

Let $a_1,...,a_{2^n} \in F$.
We can view $(a_1,...,a_{2^n})$ as a function
from $H^d$ to $F$. More precisely,
define $A: H^d \rightarrow F$ by
$A(z) = a_{\pi(z)}$.
A basic fact is that there exists a unique extension
of $A$ into a function $\tilde{A} : F^d \rightarrow F$,
such that $\tilde{A}$ is a multivariate polynomial
(in $d$ variables)
of degree at most $|H| -1$ in each variable.
The function $\tilde{A}$ is called, the {\em low degree extension}
of $a_1,...,a_{2^n}$.
Note that the total degree of $\tilde{A}$ is lower than
$2n^{1.5}/\log n < n^{1.5}$.

We define the {\em quantum low degree extension} of
$a_1,...,a_{2^n}$, by
$$|\Psi(a_1,...,a_{2^n}) \rangle = |F|^{-d/2} \cdot
 \sum_{z_1,..,z_d \in F}
 |z_1 \rangle |z_2 \rangle \cdots |z_d \rangle
 | \tilde{A}(z_1,...,z_d)\rangle.$$
Note that $|\Psi(a_1,...,a_{2^n}) \rangle$ is a quantum state
of $(d+1) c \log n = 2cn + c\log n = O(n)$ qubits.

\subsection{The Retrieval Protocol} \label{subsec:retrieval}

Assume now that Alice got the state
$|\Psi \rangle = |\Psi(a_1,...,a_{2^n}) \rangle$
and she wants to retrieve the value of $a_i$ for some $i \in [2^n]$,
or more generally, the value of $\tilde{A}(w)$
for some $w \in F^d$. This can be done by the following
interactive protocol with the infinitely powerful prover Merlin.

Alice Measures all qubits of $|\Psi \rangle$, and gets as a result
a random $z \in F^d$ and the value $\tilde{A}(z)$.
If $z \ne w$, Alice computes the line
(i.e., affine
subspace of dimension 1 in $F^d$)
that contains both $w$ and $z$.
Formally, this line is the set
$$\ell = \{w + (z-w) \cdot t \}_{t \in F} \subset F^d$$
(where all operations are in the vector space $F^d$).
Alice sends $\ell$ to Merlin\footnote{This can be done by sending
$w$ and one additional point (say, the lexicographically first point)
in $\ell$.
Note that Merlin doesn't know $z$.
(We don't care if Merlin does know $w$, but note that
we could also send $\ell$
by just sending two different points
(say, the two lexicographically first points) in it).}.
Merlin is required to respond with the value of $\tilde{A}$
on all the points in $\ell$. Denote by $g(t)$
the value given by
Merlin for the point $w + (z-w) \cdot t$.

Roughly speaking, Alice will reject (i.e., conclude the value
$Err$) if $g: F \rightarrow F$ is not a low degree polynomial in the
variable $t$, or if
$g(1)$ disagrees with the value $\tilde{A}(z)$
(which is the only value of $\tilde{A}$ that Alice knows).

Formally,
denote by $\tilde{A}|_{\ell} : F \rightarrow F$ the restriction
of  $\tilde{A}$ to the line ${\ell}$ (parameterized by $t$).
That is, $\tilde{A}|_{\ell}(t) = \tilde{A}(w + (z-w) \cdot t)$.
Recall that the total degree of $\tilde{A}$ is
$< n^{1.5}$. Hence, $\tilde{A}|_{\ell}$ is
a polynomial (in the one variable $t$) of degree
$< n^{1.5}$.
If $g$ is not a polynomial
of degree  $< n^{1.5}$ then Alice rejects automatically.
Otherwise, Alice checks whether or not $g(1) = \tilde{A}(z)$.
If
$g(1) \ne \tilde{A}(z)$ Alice rejects
(note that $g(1)$ is the value given by Merlin for the point $z$).
Otherwise, Alice concludes the value $g(0)$ (i.e., the value
given by Merlin for the point $w$).

\subsection{Analysis of the Protocol}

The analysis of the retrieval protocol of
Subsection~\ref{subsec:retrieval}
is extremely simple.

Denote by $r$ a strategy of  Merlin
in the protocol. Formally, $r$ is just the set of all
answers given by Merlin for all possible pairs $(w,\ell)$.
W.l.o.g., we can assume that the strategy $r$ is deterministic.
Denote by $V_{R1}(|\Psi \rangle, w, r)$ the value
concluded by Alice when applying the protocol on a quantum state
$|\Psi \rangle$ and a point $w \in F^d$,
when Merlin is applying the strategy~$r$.
Note that $V_{R1}(|\Psi \rangle, w, r)$ is a random
variable.
Recall that
for $a_1,...,a_{2^n} \in F$, we
denote by $\tilde{A} : F^d \rightarrow F$ the low degree
extension of $a_1,...,a_{2^n}$ and by
$|\Psi(a_1,...,a_{2^n}) \rangle$ the quantum low degree extension
of $a_1,...,a_{2^n}$,
as defined in Subsection~\ref{subsec:qlde}.

The completeness and soundness properties of the protocol
are given by the following lemma.

\begin{lemma} \label{lemma:R1}
For every $a_1,...,a_{2^n} \in F$ and every $w \in F^d$,
\begin{enumerate}
\item
$\exists r$, s.t.
$V_{R1}(|\Psi (a_1,...,a_{2^n}) \rangle, w, r) = \tilde{A}(w)$
with probability 1.
\item
$\forall r$,
$V_{R1}(|\Psi (a_1,...,a_{2^n}) \rangle, w, r)
\in \{ \tilde{A}(w) , Err \}$
with probability $\geq  1- 1/n^{c-1.5}$.
\end{enumerate}
\end{lemma}

\proof
Obviously, if Merlin's answer on line $\ell$ is
the polynomial
$g = \tilde{A}|_{\ell}$ then Alice concludes
the correct value $\tilde{A}(w)$ with probability 1.
So, the first part is obvious.

For the second part, note that if Merlin's answer on a line $\ell$
is a polynomial $g$ of degree less than $n^{1.5}$ then either $g$
is the same polynomial as $\tilde{A}|_{\ell}$ or the two polynomials
agree on less than $n^{1.5}$ points.
In the first case, Alice concludes the correct
value $\tilde{A}(w)$. In the second case, Alice will reject for every
value $z \in \ell \setminus \{ w\}$
on which the two polynomials disagree.
(Recall that Merlin doesn't know $z$ and only knows the
description of the line $\ell$).
Thus, Alice rejects on a fraction of at least $1-n^{1.5}/|F|$
of the points in $\ell \setminus \{ w\}$.
Summing over all lines, with probability of at least
$1-n^{1.5}/|F| = 1- 1/n^{c-1.5}$
Alice will either conclude the correct value
or reject.
\QED

\subsection{Retrieving More Values}

Suppose now that we want Alice to be able to retrieve
the values of $a_{i_1},...,a_{i_k}$, for $k > 1$.
An obvious way to do that is by encoding
$a_1,...,a_{2^n}$ by the tensor product of
$|\Psi(a_1,...,a_{2^n}) \rangle$ with
itself $k$ times, that is, by the state
$|\Psi \rangle \otimes \cdots \otimes |\Psi \rangle$,
where
$|\Psi \rangle = |\Psi(a_1,...,a_{2^n}) \rangle$ is the quantum
low degree extension of $a_1,...,a_{2^n}$, (as before).
Alice can now retrieve one value from each copy of
the state $|\Psi \rangle$.
Moreover, this can be done in parallel in one
round\footnote{It is not hard to show that in this setting
applying the protocol in parallel is practically equivalent
to applying it sequentially.
Issues
of parallel repetition, such as the the ones in~\cite{Raz},
are not a problem here.}.

In this paper, we will not use this method.
We will need, however, a method to retrieve more than
one value from only {\bf one} copy of
$|\Psi(a_1,...,a_{2^n}) \rangle$.
This can be done by a generalization of the retrieval protocol
of Subsection~\ref{subsec:retrieval}.
For simplicity of the presentation, we will present here the
retrieval of only two values. The same protocol generalizes
to an arbitrary~$k$.
The complexity of the retrieval protocol, however,
is exponential in $k$.
%Note that in this paper we will
%only need to use the case $k=2$.

Assume that Alice got the state
$|\Psi \rangle = |\Psi(a_1,...,a_{2^n}) \rangle$
and she wants to retrieve the values of $a_i,a_{i'}$
for some (different) $i,i' \in [2^n]$,
or more generally, the values of $\tilde{A}(w),\tilde{A}(w')$
for some (different) $w,w' \in F^d$. This can be done by the following
interactive protocol.

Alice Measures all qubits of $|\Psi \rangle$, and gets as a result
a random $z \in F^d$ and the value $\tilde{A}(z)$.
Alice computes the plane
(i.e., affine
subspace of dimension 2 in $F^d$)
that contains all three points $w,w',z$.
Formally, this plane\footnote{Note
that if $z,w,w'$ happen to be on the same line then $p$
is a line rather than a plan. Nevertheless, we can proceed
in the exact same way.}  is the set
$$p = \{w + (z-w) \cdot t_1 +
(w'-w) \cdot t_2\}_{t_1,t_2 \in F} \subset F^d.$$
Alice sends $p$ to Merlin, who
is required to respond with the value of $\tilde{A}$
on all the points in $p$. Denote by $g(t_1,t_2)$
the value given by
Merlin for the point $w + (z-w) \cdot t_1 + (w'-w) \cdot t_2$.

If $g$ is not a polynomial
of total degree  $< n^{1.5}$ then Alice rejects automatically.
If
$g(1,0) \ne \tilde{A}(z)$ Alice rejects as well.
Otherwise, Alice concludes the values $(g(0,0),g(0,1))$
(i.e., the values
given by Merlin for the points $w,w'$).

Denote by $r$ a strategy of  Merlin
in the protocol.
Denote by $V_{R2}(|\Psi \rangle, (w,w'), r)$ the values
concluded by Alice when applying the protocol on a quantum state
$|\Psi \rangle$ and points $w,w' \in F^d$,
when Merlin is applying the strategy~$r$.
The completeness and soundness properties of the protocol
are given by the following lemma.

\begin{lemma}
For every $a_1,...,a_{2^n} \in F$ and every $w,w' \in F^d$,
\begin{enumerate}
\item
$\exists r$, s.t.
$V_{R2}(|\Psi (a_1,...,a_{2^n}) \rangle, (w,w'), r) =
(\tilde{A}(w),\tilde{A}(w'))$
with probability 1.
\item
$\forall r$,
$V_{R2}(|\Psi (a_1,...,a_{2^n}) \rangle, (w,w'), r) \in
\{ (\tilde{A}(w),\tilde{A}(w')) , Err \}$
with probability $\geq  1- 1/n^{c-1.5}$.
\end{enumerate}
\end{lemma}

\proof
Same as the proof of Lemma~\ref{lemma:R1}
\QED

\section{Interactive Proofs with Quantum Advice}

In this section, we present the results discussed
in Subsection~\ref{subsec:QAM}.
Quantum interactive proof systems were first introduced
by Watrous (see~\cite{Wat} for the formal definition).
In these proof systems, the verifier can be modelled by
a polynomial-size quantum circuit.
Quantum interactive proof systems with polynomial-size
quantum advice are defined in the same way,
except that the verifier is modelled by
a polynomial-size quantum circuit
with a polynomial-size quantum advice.
That is, the verifier is a polynomial-size quantum circuit,
with working space initiated with an arbitrary
quantum state $|\Psi \rangle$.
(The state $|\Psi \rangle$ is considered to be part of the description
of the circuit and it cannot depend on the inputs to the circuit).

The class $QIP/qpoly$ is defined to be the class of all languages
that have polynomial-size quantum interactive proofs
with a polynomial-size quantum advice.
We show that the class $QIP/qpoly$ contains all languages.
For any language $L$,
the membership $x \in L$
can be proved by a polynomial-size
quantum interactive proof,
with a polynomial-size quantum advice.
Moreover, the interactive proofs that we construct
for the membership $x \in L$
are of only one round,
and all messages communicated are classical\footnote{Note
that formally in the standard definition of quantum interactive proofs
the parties can only communicate between them quantum
messages. Nevertheless, since a quantum states can encode
classical messages, the model is equivalent to a model
where the parties can communicate both quantum and classical
messages. In the interactive proofs constructed here,
the parties communicate only classical messages.}.

\begin{theorem}
$QIP/qpoly$ contains all languages.
\end{theorem}

\proof
Let $L$ be any language.
For a string $i$ of length $n$ bits, define
$a_i =1$ iff $i \in L$.
We will use as a quantum advice for the verifier
the quantum low degree extension
$|\Psi(a_{0},...,a_{2^{n}-1}) \rangle$
(see Subsection~\ref{subsec:qlde}).
The proof now follows
by the retrieval protocol of
Subsection~\ref{subsec:retrieval}.
Given $x$ of length $n$ bits, the verifier uses the
retrieval protocol to retrieve
the value of $a_x$ (by an interactive protocol with the prover).
The verifier accepts iff
the value concluded by the protocol is~$1$
(and rejects if the value concluded is 0 or $Err$).
The completeness and soundness properties follow
immediately by Lemma~\ref{lemma:R1}.
More precisely, if $x \in L$ there is a strategy
for the prover that
causes the verifier
to accept (with probability 1), and on the other
hand,
if $x \not \in L$ then
no strategy
for the prover can cause the verifier to
accept with non-negligible probability.
\QED

\section{Quantum Low Degree Testing} \label{sec: LDT}

In the settings of Subsection~\ref{subsec:Holevo}
and Subsection~\ref{subsec:QAM}, a verifier
could assume that the quantum state given to him
is a correct quantum low degree extension of some
$a_1,...,a_{2^n}$
(as defined in Subsection~\ref{subsec:qlde}).
In the setting of quantum proofs, and quantum versions
of the PCP theorem, a verifier cannot assume anything
about a quantum state given to him.
A key step towards proving the results discussed in
Subsection~\ref{subsec:PCP} is a {\em quantum low degree test},
developed in this section.
Roughly speaking,
a quantum low degree test intends to check whether
a quantum state is close to a
representation of a polynomial of small
total degree.

\subsection{Classical Low Degree Tests} \label{subsec:cldt}

Roughly speaking,
a (classical) low degree test intends to check whether
a multivariate function
is close to a polynomial of small total degree.
Low degree tests and their applications have been studied in numerous
of works and have been central in the study of interactive proofs
and probabilistic checkable proofs
(see for example~\cite{BFL,FGLSS,AS1,ALMSS}).
In this paper, we will need to use the "low-error" analysis
of~\cite{RS2,AS2} for (versions of) the test presented in~\cite{RS1}.

Let $F$ be a field and let $d$ be some integer.
Let $L$ be the set of all lines in $F^d$ (i.e., the set of all
affine subspaces of dimension 1).
For every $\ell \in L$, let $g_{\ell}: \ell \rightarrow F$
be a polynomial\footnote{We assume here that
$\ell$ is presented as $\ell = \{u + (v-u)\cdot t\}_{t \in F}$
for some $u,v \in \ell$, and hence we can think of $g_{\ell}$
as a polynomial in the variable $t$. Note that the degree
of $g_{\ell}$ does not depend on the choice of $u,v$.}
of degree $r$. Denote, $G = \{g_{\ell}\}_{\ell \in L}$.

For $f,f': F^d \rightarrow F$ and for $G = \{g_{\ell}\}_{\ell \in L}$
as above,
denote
$$\Agr[f,f'] = \P_{z \in F^d}[f(z) = f'(z)],$$
$$\Agr[f,g_{\ell}] = \P_{z \in \ell}[f(z) = g_{\ell}(z)],$$
$$\Agr[f,G] = \E_{\ell \in L}\Agr[f,g_{\ell}],$$
where all probabilities and expectations are with respect to the
uniform distribution.
%If we have a probability distribution $\mu: L \rightarrow R$,
%denote
%$$\Agr_{\mu}[f,G] = \E_{\mu}\Agr[f,g_{\ell}],$$
%where the expectation is over $\ell \in L$ chosen according
%to the probability distribution $\mu$.

The Rubinfeld-Sudan test~\cite{RS1} suggests that if
$\Agr[f,G]$ is large then $f$ is close to a polynomial
of total degree $r$.
The following lemma, that menages to work
with quite small values of $\Agr[f,G]$,
was proved in~\cite{AS2}.
A similar lemma for planes, rather than lines, was proved
in~\cite{RS2} (see also~\cite{DFKRS}).
Here, we can use any of these tests.
We note that the lemmas proved in~\cite{RS2,AS2} are in fact
stronger in several ways.
We present them here in a simpler form
that will suffice for us.

\begin{lemma} (Arora-Sudan) \label{lemma:AS2}
Let $f: F^d \rightarrow F$ be any function, and let
$G = \{g_{\ell}\}_{\ell \in L}$ be such that every
$g_{\ell}: \ell \rightarrow F$
is a polynomial
of degree $r$.
Assume that
$$\Agr[f,G]  >
\frac{c \cdot r} {|F|^{\epsilon}}$$
where $c$ is a (large enough) universal constant and
$\epsilon > 0$ is a (small enough) universal constant.
Then, there exists $h: F^d \rightarrow F$ of
total degree $r$, such that,
$$\Agr[h,f], \Agr[h,G] \geq (\Agr[f,G])^2/ 32.$$
\end{lemma}

Note that the lemma shows that
if
$\Agr[f,G]$ is large then {\bf both} $f$ and $G$
are close to a polynomial $h$
of total degree $r$.
Interestingly, it will be easier for us to use the claim about
$G$.

In this paper, we will need a slightly more general version
of Lemma~\ref{lemma:AS2}, where we allow the polynomials
$g_{\ell}: \ell \rightarrow F$ to take multiple values.
More generally, we allow each $g_{\ell}$ to be a random variable,
distributed
over polynomials of degree~$r$.
We update the above notations as follows.

For $f: F^d \rightarrow F$ and for $G = \{g_{\ell}\}_{\ell \in L}$
as above,
denote
$$\Agr[f,g_{\ell}] = \E_{g_{\ell}} \P_{z \in \ell}[f(z) = g_{\ell}(z)],$$
%$$\Agr[f,g_{\ell}] = \E_{z \in \ell}\P_{g_{\ell}}[f(z) = g_{\ell}(z)],$$
$$\Agr[f,G] = \E_{\ell \in L}\Agr[f,g_{\ell}].$$

It is a folklore meta-theorem that all  known low degree tests work
as well when assignments can take multiple values.
As before, if
$\Agr[f,G]$ is large then both $f$ and $G$ are close to a polynomial
$h$ of total degree $r$.

\begin{lemma} \label{lemma:AS2+}
Let $f: F^d \rightarrow F$ be any function, and let
$G = \{g_{\ell}\}_{\ell \in L}$ be such that every
$g_{\ell}: \ell \rightarrow F$
is a random variable, distributed over polynomials
of degree $r$.
Assume that
$$\Agr[f,G]  >
\frac{c \cdot r} {|F|^{\epsilon}}$$
where $c$ is a (large enough) universal constant and
$\epsilon > 0$ is a (small enough) universal constant.
Then, there exists $h: F^d \rightarrow F$ of
total degree $r$, such that,
$$\Agr[h,f], \Agr[h,G] \geq (\Agr[f,G])^2/ 32.$$
\end{lemma}

The lemma follows by a reduction
to Lemma~\ref{lemma:AS2}, using well known methods
(see for example~\cite{AS2,RS2,DFKRS}).

\subsection{The Quantum Test} \label{subsec: qldt}

Let $F$ be a field of size $2^a$ for some integer $a$,
and let $d$ be some integer. Recall that
our basic quantum element is a quantum register of $a$ qubits,
rather than a single qubit.
Each such basic element represents a member of the
Hilbert space $C^{|F|}$.
Denote by ${\cal H}_{d+1}$ and ${\cal H}_{2}$ the
following Hilbert spaces
$${\cal H}_{d+1} = C^{|F|^{d+1}},$$
$${\cal H}_{2} = C^{|F|^{2}}.$$
Let $L$ be the set of all lines in $F^d$ (as before).
Our quantum low degree test intends to check whether
a quantum state $|\Phi \rangle \in {\cal H}_{d+1}$
is close to a state of
the form
$$|F|^{-d/2} \cdot
 \sum_{z_1,..,z_d \in F}
 |z_1 \rangle |z_2 \rangle \cdots |z_d \rangle
 | f(z_1,...,z_d)\rangle,$$
where $f: F^d \rightarrow F$ is some polynomial of total
degree $r$.
In addition to the state $|\Phi \rangle$, the test has access to
a set of {\bf (classical)} polynomials
$G = \{g_{\ell}\}_{\ell \in L}$, where as before,
for every $\ell \in L$ the polynomial $g_{\ell}: \ell \rightarrow F$
is of degree $r$
(see footnote in Subsection~\ref{subsec:cldt}).
Each $g_{\ell}$ is supposed to be
(in a correct proof) the restriction of $f$ to the line $\ell$.
In our test,
the verifier
reads only one of the polynomials $g_{\ell}$.

\subsubsection{Step I}

The verifier chooses a random
regular (i.e., one to one) linear function
$E: F^d \rightarrow F^d$.
The function $E$ defines a permutation $U_E$ over the standard
basis for ${\cal H}_{d+1}$, as follows:
For every $z=(z_1,...,z_d) \in F^d$ and every $y \in F$,
$$ |z_1 \rangle |z_2 \rangle \cdots |z_d \rangle
 | y\rangle \longmapsto
 |E(z)_1 \rangle |E(z)_2 \rangle
  \cdots |E(z)_d \rangle
 | y\rangle .$$
Since  $U_E$ is a permutation over a
basis for ${\cal H}_{d+1}$,
it extends to a unitary transformation
$$U_E : {\cal H}_{d+1} \rightarrow {\cal H}_{d+1}.$$

The verifier computes the quantum state $U_E |\Phi \rangle$
and measures the first $d-1$ registers of that state
(i.e., $|E(z)_1 \rangle \cdots |E(z)_{d-1} \rangle$).
Denote by $b_1,...,b_{d-1} \in F$ the results of the measurement.
The state $U_E |\Phi \rangle$ collapses into a state
$|\Phi' \rangle \in {\cal H}_{2}$ (in the last two registers).

Note that the set of solutions for the set of linear equations
$$E(z)_1=b_1, \ldots, E(z)_{d-1} = b_{d-1}$$ is a line $\ell \in L$.
The line $\ell$ can be presented as
$\ell = \{u + (v-u) \cdot t \}_{t \in F} $,
where $u \in F^d$ is the unique solution for the
set of linear equations
$E(u)=(b_1, \ldots, b_{d-1}, 0),$
%$$E(u)_1=b_1, \ldots, E(u)_{d-1} = b_{d-1}, E(u)_{d} = 0,$$
and
$v \in F^d$ is the unique solution for the
set of linear equations
$E(v)=(b_1, \ldots, b_{d-1}, 1).$
%$$E(v)_1=b_1, \ldots, E(v)_{d-1} = b_{d-1}, E(v)_{d} = 1.$$

If the original state
$|\Phi \rangle $
is indeed of
the form
$$|F|^{-d/2} \cdot
 \sum_{z_1,..,z_d \in F}
 |z_1 \rangle |z_2 \rangle \cdots |z_d \rangle
 | f(z_1,...,z_d)\rangle,$$
then the collapsed state $|\Phi' \rangle \in {\cal H}_{2}$ will be
$$|\Phi' \rangle = |F|^{-1/2} \cdot
 \sum_{t \in F}
 |t \rangle
 | f_{\ell}(t)\rangle,$$
where $f_{\ell} : F \rightarrow F$ is the restriction
of $f$ to the line $\ell$ (parameterized by $t$),
i.e., $f_{\ell}(t) = f(u + (v-u) \cdot t)$.

\subsubsection{Step II}

The verifier reads the polynomial $g_{\ell}$.
We can think of this polynomial as a polynomial
$g_{\ell} : F \rightarrow F$, where the line $\ell$ is
parameterized by the same $t$ as above
(i.e., the line $\ell$ is presented as
$\ell = \{u + (v-u) \cdot t \}_{t \in F} $).

Denote by $|e_1 \rangle \in {\cal H}_{2}$ the quantum state
$$|e_1 \rangle = |F|^{-1/2} \cdot
 \sum_{t \in F}
 |t \rangle
 | g_{\ell}(t)\rangle.$$
The verifier wants to compare the
states $|\Phi' \rangle$ and $|e_1 \rangle$.
This is done as follows.
The verifier extends
$|e_1 \rangle$ into any orthonormal basis
$\{ |e_1 \rangle, \ldots, |e_{|F|^2} \rangle \}$
for the space ${\cal H}_{2}$,
and measures the state $|\Phi' \rangle$ according to this basis.
The verifier accepts
if the result of the measurement is~1
and rejects in any other case.

Note that if indeed
$$|\Phi' \rangle = |F|^{-1/2} \cdot
 \sum_{t \in F}
 |t \rangle
 | f_{\ell}(t)\rangle,$$
and $f_{\ell} = g_{\ell}$, then $|\Phi' \rangle = |e_1 \rangle$
and the verifier accepts with probability 1.
In general, the verifier accepts with probability
$$|\langle e_1 | \Phi'\rangle|^2.$$

\subsection{Complexity of the Verifier}

The complexity of the verifier in the procedure of
Subsection~\ref{subsec: qldt} is polynomial
in $|F|$ and $d$. To see this, we need to check that both steps
can be done in that complexity.

In the first step, the verifier needs to compute the quantum
transformation
$$ |z_1 \rangle |z_2 \rangle \cdots |z_d \rangle
  \longmapsto
 |E(z)_1 \rangle |E(z)_2 \rangle
  \cdots |E(z)_d \rangle.$$
It is enough to show that the classical transformation
$$ (z_1,z_2,\ldots ,z_d)
  \longmapsto
 (E(z)_1, E(z)_2, \ldots,
E(z)_d)$$
has a reversible classical circuit of size $poly(|F|,d)$.
This follows immediately by the fact that any such transformation
can be expressed as a product of $poly(d)$ reversible operations
on only two variables each. One way to do that is by the
inverse of the Gauss elimination procedure, that
shows how to diagonalize any $d \times d$ matrix $E$ by
a sequence of $poly(d)$ operations that work on only two rows
each.
Note that every operation that works on only two variables
can be trivially translated into a quantum circuit
of size $poly(|F|)$, as the dimension of the relevant Hilbert
space, ${\cal H}_{2}$, is $|F|^2$.

In the second step, the verifier needs to
measure the
state $|\Phi' \rangle$ according to the basis
$\{ |e_1 \rangle, \ldots, |e_{|F|^2} \rangle \}$.
Note however that since
the space ${\cal H}_{2}$ is of dimension $|F|^2$,
this can trivially be done by a quantum circuit of size
$poly(|F|)$.

\subsection{Analysis of the Test}

For a quantum state
$|\Phi \rangle \in {\cal H}_{d+1}$
and for
a set of polynomials
$G = \{g_{\ell}\}_{\ell \in L}$ (where
for every $\ell \in L$ the polynomial $g_{\ell}: \ell \rightarrow F$
is of degree $r$),
denote by $V_{QLDT}(|\Phi \rangle,G)$
the probability that the quantum low degree test procedure
of Subsection~\ref{subsec: qldt} accepts.

The completeness of the test
is given by the following
lemma.
The lemma shows that if $|\Phi \rangle$ is indeed a correct
representation of a polynomial $f: F^d \rightarrow F$
of total degree $r$, and each $g_{\ell}$ is the restriction
of $f$ to the line $\ell$, then the test accepts
with probability~1.

\begin{lemma}
Assume that
$$|\Phi \rangle = |F|^{-d/2} \cdot
 \sum_{z_1,..,z_d \in F}
 |z_1 \rangle |z_2 \rangle \cdots |z_d \rangle
 | f(z_1,...,z_d)\rangle,$$
for some polynomial
$f: F^d \rightarrow F$ of total degree $r$.
Assume that
$G = \{g_{\ell}\}_{\ell \in L}$, where
%for every $\ell \in L$ the polynomial $g_{\ell}: \ell \rightarrow F$
every $g_{\ell}: \ell \rightarrow F$
is the restriction of $f$ to the line $\ell$.
Then,
$$V_{QLDT}(|\Phi \rangle,G)=1.$$
\end{lemma}

\proof
The proof is straightforward.
As mentioned above, after Step I we get
the collapsed state $|\Phi' \rangle= |e_1 \rangle$.
Hence, the result of the measurement
in Step~II will always be 1.
\QED

The soundness of the test
is harder to prove and is given by the following
lemma.
The lemma shows that if $V_{QLDT}(|\Phi \rangle,G)$
is large then $G$ is close to
a polynomial $h$ of total degree $r$.
Recall that the original motivation of the test was to prove that
$|\Phi \rangle$ is close to a representation of
a polynomial $h$ of low total degree.
Nevertheless, it will be enough for us to have this
property
for $G$
rather than $|\Phi \rangle$.
For simplicity of the presentation,
we state and prove the lemma only for $G$.

%\begin{lemma} \label{lemma:qldt-sound}
%Let $|\Phi \rangle \in {\cal H}_{d+1}$ be any
%quantum state, and let
%$G = \{g_{\ell}\}_{\ell \in L}$ be such that every
%$g_{\ell}: \ell \rightarrow F$
%is a polynomial
%of degree $r$.
%Assume that
%$$V_{QLDT}(|\Phi \rangle,G)  >
%\frac{c \cdot r} {|F|^{\epsilon}}$$
%where $c$ is a (large enough) universal constant and
%$\epsilon > 0$ is a (small enough) universal constant.
%Then, there exists $h: F^d \rightarrow F$ of
%total degree $r$, such that,
%$$\Agr[h,G] \geq (V_{QLDT}(|\Phi \rangle,G))^{c'}/ c',$$
%where $c'$ is a (large enough) universal constant.
%\end{lemma}

\begin{lemma} \label{lemma:qldt-sound}
Let
$G = \{g_{\ell}\}_{\ell \in L}$ be such that every
$g_{\ell}: \ell \rightarrow F$
is a polynomial
of degree $r$.
Assume that for some quantum state $|\Phi \rangle \in {\cal H}_{d+1}$,
$$V_{QLDT}(|\Phi \rangle,G)  >
\frac{c \cdot r} {|F|^{\epsilon}}$$
where $c$ is a (large enough) universal constant and
$\epsilon > 0$ is a (small enough) universal constant.
Then, there exists $h: F^d \rightarrow F$ of
total degree $r$, such that,
$$\Agr[h,G] \geq [V_{QLDT}(|\Phi \rangle,G)]^{4}/ 50.$$
\end{lemma}

The proof of the Lemma is given in
Subsection~\ref{subsect:qldt-sound}.

As in the case of Lemma~\ref{lemma:AS2}, we
will need a slightly more general version
of Lemma~\ref{lemma:qldt-sound}, where we allow the polynomials
$g_{\ell}: \ell \rightarrow F$ to take multiple values.
More generally, we allow each $g_{\ell}$ to be a random variable,
distributed
over polynomials of degree~$r$.
When reading $g_{\ell}$, the verifier gets an
evaluation of $g_{\ell}$, that is, each polynomial
of degree~$r$ is obtained with the probability
that $g_{\ell}$ gets that value.

We
denote by $V_{QLDT}(|\Phi \rangle,G)$
the probability that the quantum low degree test procedure
accepts on
a quantum state
$|\Phi \rangle$
and on a set
$G = \{g_{\ell}\}_{\ell \in L}$ as above.

%\begin{lemma} \label{lemma:qldt-sound+}
%Let $|\Phi \rangle \in {\cal H}_{d+1}$ be any
%quantum state, and let
%$G = \{g_{\ell}\}_{\ell \in L}$ be such that
%every
%$g_{\ell}: \ell \rightarrow F$
%is a random variable, distributed over polynomials
%of degree $r$.
%Assume that
%$$V_{QLDT}(|\Phi \rangle,G)  >
%\frac{c \cdot r} {|F|^{\epsilon}}$$
%where $c$ is a (large enough) universal constant and
%$\epsilon > 0$ is a (small enough) universal constant.
%Then, there exists $h: F^d \rightarrow F$ of
%total degree $r$, such that,
%$$\Agr[h,G] \geq (V_{QLDT}(|\Phi \rangle,G))^{c'}/ c',$$
%where $c'$ is a (large enough) universal constant.
%\end{lemma}

\begin{lemma} \label{lemma:qldt-sound+}
Let
$G = \{g_{\ell}\}_{\ell \in L}$ be such that
every
$g_{\ell}: \ell \rightarrow F$
is a random variable, distributed over polynomials
of degree $r$.
Assume that for some quantum state $|\Phi \rangle \in {\cal H}_{d+1}$,
$$V_{QLDT}(|\Phi \rangle,G)  >
\frac{c \cdot r} {|F|^{\epsilon}}$$
where $c$ is a (large enough) universal constant and
$\epsilon > 0$ is a (small enough) universal constant.
Then, there exists $h: F^d \rightarrow F$ of
total degree $r$, such that,
$$\Agr[h,G] \geq [V_{QLDT}(|\Phi \rangle,G)]^{4}/ 50.$$
\end{lemma}

The proof of Lemma~\ref{lemma:qldt-sound+} is the same
as the one of Lemma~\ref{lemma:qldt-sound}, using
Lemma~\ref{lemma:AS2+} rather than Lemma~\ref{lemma:AS2}.

\subsection{Proof of Lemma~\ref{lemma:qldt-sound}}
\label{subsect:qldt-sound}

\subsubsection{Notations}
First note that w.l.o.g. we can assume that $|\Phi \rangle $
is a {\bf pure state}.
For $z=(z_1,...,z_d) \in F^d$ and $y \in F$, denote by
$\phi_{z,y}$ the coefficient of
$ |z_1 \rangle |z_2 \rangle \cdots |z_d \rangle
 | y\rangle$ in  $|\Phi \rangle $. That is,
$$|\Phi \rangle =
 \sum_{z \in F^d, y \in F}
 \phi_{z,y}
 |z \rangle
| y\rangle.$$
For every $z \in F^d$, denote
$$ \phi_{z} = \sqrt{\sum_{y \in F} |\phi_{z,y}|^2}$$
For every line $\ell \in L$, denote
$$ \phi_{\ell} = \sqrt{\sum_{z \in \ell} |\phi_{z}|^2}$$
Denote
$$N=\frac{|F|^d-1}{|F|-1}$$
and note that $N$ is the number of directions of lines
$\ell$ in $F^d$.
For every $z \in F^d$, denote by $L(z) \subset L$ the set
of lines $\ell$ that contain $z$. Note that  every
$L(z)$ is a set of $N$ lines.

We denote the total acceptance probability
$V_{QLDT}(|\Phi \rangle,G)$ by $\gamma$, i.e.,
$$\gamma = V_{QLDT}(|\Phi \rangle,G).$$

\subsubsection{The Acceptance Probability}

Assume w.l.o.g. that every $\phi_{z,y}$
(and hence also every $\phi_{z}$ and every $\phi_{\ell}$)
is none-zero. Otherwise, we just change the state $|\Phi \rangle$
to an extremely close state that satisfies that property.
(This is done for the simplicity of the presentation,
in order to avoid divisions by 0).

In Step I of the test, the linear function $E$ determines
the direction of the line $\ell$ that we obtain in that step.
Since $E$ is chosen with the uniform distribution,
each direction is chosen with probability $1/N$.
After the direction is chosen, each line $\ell$ in that direction
is obtained with probability $\phi_{\ell}^2$.
Altogether, every line $\ell$ is obtained with
probability $\phi_{\ell}^2/N$.

If a line $\ell$ was obtained, the state
$|\Phi \rangle$ collapses to the state
$$|\Phi'_{\ell} \rangle = \phi_{\ell}^{-1} \cdot
 \sum_{t \in F} \sum_{y \in F}
 \phi_{z(t),y} |t \rangle
| y \rangle,$$
where $z(t) = u + (v-u) \cdot t$, and $u,v$ are
the ones defined in Subsection~\ref{subsec: qldt}
(i.e., $u,v$ are such that
the line $\ell$ is presented as
$\ell = \{u + (v-u) \cdot t \}_{t \in F} $,
as described in Subsection~\ref{subsec: qldt}).

Since
$|e_1 \rangle = |F|^{-1/2} \cdot
 \sum_{t \in F}
 |t \rangle
 | g_{\ell}(t)\rangle$,
the acceptance probability
$|\langle e_1 | \Phi'_{\ell}\rangle|^2$ is
\begin{eqnarray*}
%\label{eqn: leq1}
|\langle e_1 | \Phi'_{\ell}\rangle|^2 =
\left|  \phi_{\ell}^{-1} \cdot |F|^{-1/2} \cdot
 \sum_{t \in F}
 \phi_{z(t),g_{\ell}(t)} \right|^2 =
\left|  \phi_{\ell}^{-1} \cdot |F|^{-1/2} \cdot
 \sum_{z \in \ell}
 \phi_{z,g_{\ell}(z)} \right|^2
\end{eqnarray*}
 where (for simplicity) we think of $g_{\ell}$ as a polynomial
$g_{\ell}: F \rightarrow F$ when we write $g_{\ell}(t)$,
and as a polynomial
$g_{\ell}: \ell \rightarrow F$ when we write $g_{\ell}(z)$.

The total acceptance probability $\gamma$
can now be expressed
as
%\begin{eqnarray}
%\gamma =
%\sum_{\ell \in L} (\phi_{\ell}^2/N) \cdot
%|\langle e_1 | \Phi'_{\ell}\rangle|^2
%\end{eqnarray}
\begin{eqnarray*}
%\label{eqn:gamma-eq}
\gamma =
\sum_{\ell \in L} \left( \frac{\phi_{\ell}^2}{N} \right) \cdot
\left|  \phi_{\ell}^{-1} \cdot |F|^{-1/2} \cdot
\sum_{z \in \ell}
\phi_{z,g_{\ell}(z)} \right|^2
=
(|F| \cdot N)^{-1} \cdot
\sum_{\ell \in L}
\left|
\sum_{z \in \ell}
\phi_{z,g_{\ell}(z)} \right|^2
\end{eqnarray*}
We can see from this expression that
w.l.o.g. we can assume that all the
coefficients $\phi_{z,y}$ of the state $|\Phi \rangle$
are real and positive.
(Otherwise, we change each $\phi_{z,y}$ to $|\phi_{z,y}|$
and we can only increase the total acceptance probability).
Hence,
\begin{eqnarray*}
\gamma =
(|F| \cdot N)^{-1} \cdot
\sum_{\ell \in L}
\left(
\sum_{z \in \ell}
\phi_{z,g_{\ell}(z)} \right) \cdot
\left(
\sum_{z' \in \ell}
\phi_{z',g_{\ell}(z')} \right)
\end{eqnarray*}
\begin{eqnarray*}
=
(|F| \cdot N)^{-1} \cdot
\sum_{\ell \in L}
\sum_{z,z' \in \ell}
\phi_{z,g_{\ell}(z)}  \cdot
\phi_{z',g_{\ell}(z')}
\end{eqnarray*}
Since every $\phi_{z',g_{\ell}(z')}$ is at most
$\phi_{z'}$, we can bound
\begin{eqnarray*}
\gamma \leq
(|F| \cdot N)^{-1} \cdot
\sum_{\ell \in L}
\sum_{z,z' \in \ell}
\phi_{z,g_{\ell}(z)}  \cdot
\phi_{z'}
\end{eqnarray*}
We will write the last expression as a sum of two expressions,
according to whether or not $z=z'$.
The first expression is the sum of all terms where $z=z'$.
That expression is
\begin{eqnarray*}
(|F| \cdot N)^{-1} \cdot
\sum_{\ell \in L}
\sum_{z \in \ell}
\phi_{z,g_{\ell}(z)}  \cdot
\phi_{z}
\leq
(|F| \cdot N)^{-1} \cdot
\sum_{\ell \in L}
\sum_{z \in \ell}
\phi_{z}^2
\end{eqnarray*}
\begin{eqnarray*}
=
(|F| \cdot N)^{-1} \cdot
\sum_{z \in F^d}
\sum_{\ell \in L(z)}
\phi_{z}^2
=
|F|^{-1} \cdot
\sum_{z \in F^d}
\phi_{z}^2
=
|F|^{-1}
\end{eqnarray*}
Hence,
\begin{eqnarray*}
\gamma - |F|^{-1} \leq
(|F| \cdot N)^{-1} \cdot
\sum_{\ell \in L}
\sum_{z \ne z' \in \ell}
\phi_{z,g_{\ell}(z)}  \cdot
\phi_{z'}
\end{eqnarray*}
\begin{eqnarray*}
=
(|F| \cdot N)^{-1} \cdot
\sum_{\ell \in L}
\sum_{z \ne z' \in \ell}
\left( \phi_{z}  \cdot
\phi_{z'} \right)
\cdot
(\phi_{z,g_{\ell}(z)} / \phi_{z} )
\end{eqnarray*}
and hence by the Cauchy-Schwartz inequality,
\begin{eqnarray*}
\gamma - |F|^{-1} \leq
(|F| \cdot N)^{-1} \cdot
\sqrt{\sum_{\ell \in L}
\sum_{z \ne z' \in \ell}
\left( \phi_{z}  \cdot
\phi_{z'} \right)^2}
\cdot
\sqrt{
\sum_{\ell \in L}
\sum_{z \ne z' \in \ell}
(\phi_{z,g_{\ell}(z)} / \phi_{z} )^2}
\end{eqnarray*}
In this formula, we can substitute
\begin{eqnarray*}
\sum_{\ell \in L}
\sum_{z \ne z' \in \ell}
\left( \phi_{z}  \cdot
\phi_{z'} \right)^2
=
\sum_{z \in F^d}
\sum_{z' \ne z \in F^d}
\left( \phi_{z}  \cdot
\phi_{z'} \right)^2
\leq
\sum_{z \in F^d}
\sum_{z' \in F^d}
\left( \phi_{z}  \cdot
\phi_{z'} \right)^2
=
\left(
\sum_{z \in F^d}
\phi_{z}^2   \right)^2 =1
\end{eqnarray*}
and
\begin{eqnarray*}
\sum_{\ell \in L}
\sum_{z \ne z' \in \ell}
(\phi_{z,g_{\ell}(z)} / \phi_{z} )^2
=
\sum_{z \in F^d}
\sum_{\ell \in L(z)}
\sum_{z' \ne z \in \ell}
(\phi_{z,g_{\ell}(z)} / \phi_{z} )^2
=
(|F|-1) \cdot
\sum_{z \in F^d}
\sum_{\ell \in L(z)}
(\phi_{z,g_{\ell}(z)} / \phi_{z} )^2
\end{eqnarray*}
and we get
\begin{eqnarray*}
\left( \gamma - |F|^{-1} \right)^2 \leq
(|F| \cdot N)^{-2} \cdot
(|F|-1) \cdot
\sum_{z \in F^d}
\sum_{\ell \in L(z)}
(\phi_{z,g_{\ell}(z)} / \phi_{z} )^2
\end{eqnarray*}
\begin{eqnarray} \label{ineq:agg}
\leq
|F|^{-d} \cdot N^{-1} \cdot
\sum_{z \in F^d}
\sum_{\ell \in L(z)}
(\phi_{z,g_{\ell}(z)} / \phi_{z} )^2.
\end{eqnarray}

\subsubsection{Using the Classical Test}

We will now define a probabilistic function
$f: F^d \rightarrow F$.
Formally, for every $z \in F^d$ we define $f(z)$
as a random variable in $F$. Alternatively, we can think
of $f$ as a distribution over functions from $F^d$ to $F$.
For every $z \in F^d$ and $y \in F$, we define
$$\P[f(z)=y]
= \left( \phi_{z,y} / \phi_{z} \right)^2.$$
Note that
$$\sum_{y \in F} ( \phi_{z,y} / \phi_{z} )^2 =1.$$

We extend the definition of $\Agr[f,G]$ from
Subsection~\ref{subsec:cldt} to probabilistic functions~$f$.
Formally, we define
$$\Agr[f,g_{\ell}] = \E_{f}
\P_{z \in \ell}[f(z) = g_{\ell}(z)],$$
$$\Agr[f,G] = \E_{\ell \in L}\Agr[f,g_{\ell}].$$

Thus,
\begin{eqnarray*}
\Agr[f,G] =
\E_{\ell \in L} \E_{f} \P_{z \in \ell}
[f(z) = g_{\ell}(z)]
\end{eqnarray*}
\begin{eqnarray*}
=
\E_{\ell \in L} \E_{z \in \ell} \P_{f}
[f(z) = g_{\ell}(z)]
\end{eqnarray*}
\begin{eqnarray*}
=
\E_{z \in F^d} \E_{\ell \in L(z)} \P_{f}
[f(z) = g_{\ell}(z)]
\end{eqnarray*}
\begin{eqnarray*}
=
|F|^{-d} \cdot N^{-1} \cdot
\sum_{z \in F^d}
\sum_{\ell \in L(z)}
 \P_{f}
[f(z) = g_{\ell}(z)]
\end{eqnarray*}
\begin{eqnarray*}
=
|F|^{-d} \cdot N^{-1} \cdot
\sum_{z \in F^d}
\sum_{\ell \in L(z)}
(\phi_{z,g_{\ell}(z)} / \phi_{z} )^2
\end{eqnarray*}
\begin{eqnarray*}
\geq
\left( \gamma - |F|^{-1} \right)^2
\end{eqnarray*}
(by inequality~\ref{ineq:agg}).

Since we can think of $f$ as a distribution over
deterministic functions $f'$, the agreement
$\Agr[f,G]$ is a convex combination of
$\Agr[f',G]$ for deterministic functions $f'$.
Hence, there exists a (deterministic)
function $f': F^d \rightarrow F$ with
\begin{eqnarray*}
\Agr[f',G]
\geq
\left( \gamma - |F|^{-1} \right)^2.
\end{eqnarray*}
Hence, by Lemma~\ref{lemma:AS2},
there exists $h: F^d \rightarrow F$ of
total degree $r$, such that,
$$\Agr[h,G] \geq \gamma^4/ 50$$
(under the assumption that the universal constant $c$ is large
enough and the universal constant $\epsilon$ is small enough).
\QED

\section{Quantum Information and the PCP Theorem}

In this section, we present the results discussed
in Subsection~\ref{subsec:PCP}.
Roughly speaking,
we show that
the membership $x \in SAT$
can be proved by a logarithmic-size
quantum state, together with a
polynomial-size classical proof of
blocks of poly-logarithmic length,
such that after measuring the quantum state
the verifier only needs to read one of the blocks
of the classical proof.

\subsection{Quantum PCP}

In all that comes below, a verifier is a polynomial-time
machine that can process both quantum states
and classical strings.

We define an $(s_1,s_2)$-verifier to be as follows:
The verifier gets
three inputs: $(x,| \Phi \rangle, p)$.
The first input,
$x$, is a classical string of length $n$ bits.
(We think of $x$ as the input whose membership in a language
$L$ is to be verified).
The second input, $|\Phi \rangle$, is a quantum state
of length $s_1$ qubits.
The third input, $p$, is a classical array
of $poly(n)$ blocks of length
$s_2$ bits each.
(We think of $(| \Phi \rangle, p)$ as a possible proof for the
membership $x \in L$).
The verifier is allowed to query at most {\bf one} of the blocks
of the third input $p$.

We define the class
$\QPCP[s_1,s_2,\epsilon]$ as follows:
A language $L$ is in $\QPCP[s_1,s_2,\epsilon]$ if there exists
an $(O(s_1),O(s_2))$-verifier $V$, such that the following
completeness and soundness
properties are satisfied:
\begin{enumerate}
\item
For any $x \in L$, there
exist $|\Phi \rangle$ and $p$, such that
$$\P[V(x,|\Phi \rangle,p) = accept] =1.$$
\item
For any $x \not \in L$, and any $|\Phi \rangle$ and $p$,
$$\P[V(x,|\Phi \rangle,p) = accept] \leq \epsilon.$$
\end{enumerate}
The definition extends to promise problems, where we only consider
inputs $x$ that satisfy a certain {\em promise}.

\begin{theorem} \label{theorem:qpcp}
$SAT \in \QPCP[log (n),polylog(n),o(1)]$
\end{theorem}

\subsection{Proof of Theorem~\ref{theorem:qpcp}}

\subsubsection{Classical PCP} \label{subsubsec: cpcp}

For the proof of Theorem~\ref{theorem:qpcp},
it is clearly enough to prove
$L \in \QPCP[log (n),polylog(n),o(1)]$ for any other
$NP$-complete language
or promise problem $L$.
We will work with the following promise problem
that we call $GAP(s,q,\epsilon)$:

An instance of the problem is
$x=(\varphi_1,...,\varphi_k)$, where
$\varphi_1,...,\varphi_k$ are predicates
over a set of variables $\{Y_1,...,Y_m\}$.
Every variable $Y_i$ can take $2^s$ different values
(i.e., we can think of every $Y_i$ as a block of $s$ bits).
Every predicate $\varphi_i$ depends on at most $q$
of the variables $Y_1,...,Y_m$.
The promise is that: either, there is an assignment to
$Y_1,...,Y_m$ that satisfies {\bf all} predicates,
or, any assignment to $Y_1,...,Y_m$ satisfies
at most $\epsilon$ fraction of the predicates.
The goal is to accept iff the first possibility is correct
(under the assumption that the promise is satisfied).

Different versions of the PCP theorem prove the $NP$-completeness
of $GAP(s,q,\epsilon)$ for a large range of values of the
parameters $s,q,\epsilon$.
Here, we will be interested in the following parameters:
We require $s$ to be at most $\log \log n$
(where $n$ is the length of $x$).
We require $q$ to be constant.
We would like $\epsilon$ to be as small as possible,
preferably sub-constant.
(The error of our verifier in the proof of
Theorem~\ref{theorem:qpcp}
will be polynomially
related to $\epsilon$. Thus, a small constant $\epsilon$
is ok if we only want to achieve a small constant error).

The $NP$-completeness of
$GAP(s,q,\epsilon)$, for some
$s \leq \log \log n$, for some constant $q$, and for some
$\epsilon \leq (\log n)^{-\Omega(1)}$ is
known~\cite{RS2,AS2,DFKRS}.
Moreover, if we only tried to achieve
a small {\bf constant} probability
of error, we could have used many other versions of the PCP theorem.
For example, we could have used the results in~\cite{Raz} and work with
$q=2$ and an arbitrarily small constant $\epsilon$.

In all that comes below, we fix $(s,q,\epsilon)$ such that
$GAP(s,q,\epsilon)$ is known to be $NP$-complete and such that:
$q$ is constant, $s$ is at most $\log \log n$, and $\epsilon$
is sub-constant.
We will show that the problem is
in $\QPCP[log (n),polylog(n),o(1)]$.
The best probability of error that we are able to achieve is
$(\log n)^{-\Omega(1)}$.

We will construct a verifier $V$ such that on an instance
$x=(\varphi_1,...,\varphi_k)$
of $GAP(s,q,\epsilon)$, the following properties are satisfied:
\begin{enumerate}
\item
If there exists an assignment to
$Y_1,...,Y_m$ that satisfies all predicates
$\varphi_1,...,\varphi_k$,
then there
exist $|\Phi \rangle$ and $p$, such that
$$\P[V(x,|\Phi \rangle,p) = accept] =1.$$
\item
If any assignment to $Y_1,...,Y_m$ satisfies
at most $\epsilon$ fraction of the predicates
$\varphi_1,...,\varphi_k$, then for
any $|\Phi \rangle$ and $p$,
$$\P[V(x,|\Phi \rangle,p) = accept] \leq \epsilon'$$
(where $\epsilon' = o(1)$).
\end{enumerate}

Every assignment to $Y_1,...,Y_m$ that satisfies all predicates
will translate into $(|\Phi \rangle, p)$, such that,
$\P[V(x,|\Phi \rangle,p) = accept] =1$.
We think of $(|\Phi \rangle, p)$ as a proof for
the satisfiability of $\varphi_1,...,\varphi_k$.
We think of the verifier as a procedure that verifies
that proof.
We will first describe how to translate an assignment
to $Y_1,...,Y_m$ (that satisfies all predicates)
into a (correct) proof $(|\Phi \rangle, p)$,
and then describe the verification procedure.

\subsubsection{Preliminaries}

Let
$x=(\varphi_1,...,\varphi_k)$ be an instance of
length $n$ of $GAP(s,q,\epsilon)$.
For simplicity of the notations,
extend the set of variables $\{Y_1,...,Y_m\}$
to $\{Y_1,...,Y_n\}$, (by adding dummy variables).
Denote by $t_1,...,t_k$ the sets of variables that
the predicates
$\varphi_1,...,\varphi_k$ depend on (respectively).
%and denote $T= \{t_1,...,t_k\}$.
Recall that each $t_j$ is a set of size at most $q$.
W.l.o.g., we can assume that $t_1,...,t_k$ are all
different. (This is only done for simplifying the notations).
%W.l.o.g., we can assume that each $t_i$ is a set
%of size {\bf exactly} $q$.
W.l.o.g., we can assume that every predicate in
$\varphi_1,...,\varphi_k$ has at least one satisfying
assignment. (Otherwise, it is clear that there is no
assignment that satisfies all predicates, and since $s$
is at most $\log \log n$ the verifier can check that easily).

W.l.o.g., assume that $n = 2^{\hat{n}}$, such that
$\hat{n} > 4$ is an even power of 2
(otherwise, we just increase $n$ to at most $n^4$).
The variable $\hat{n}$ will play the role of $n$ in
Subsection~\ref{subsec:qlde}.
As in Subsection~\ref{subsec:qlde},
denote by $F$ a field of size $2^a \doteq \hat{n}^c$, where $c$
is a large enough constant integer that will be determined later on.
As in Subsection~\ref{subsec:qlde},
denote $d = 2\hat{n}/\log \hat{n}$ and assume for simplicity
that $d$ is integer.
Assume that $d > q+1$.
Note that $|F^d|$ is polynomial in $n$.

Let $H \subset F$ be a
subset of size $\sqrt{\hat{n}}$,
as in Subsection~\ref{subsec:qlde}, and let
$\pi : H^d \rightarrow [2^{\hat{n}}] $  be a
one-to-one function, as in Subsection~\ref{subsec:qlde}.
We will use here the inverse function
$\pi^{-1} : [n] \rightarrow F^d $.
This function maps the variables in $\{Y_1,...,Y_n\}$
to $F^d$.
Intuitively, we think of each variable $Y_i$ as placed
on the point $\pi^{-1}(i) \in F^d$.
For every $t_j$, define
$\tau_j = \pi^{-1}(t_j) \subset F^d$.
%Define ${\cal T} = \{\tau_1,...,\tau_k\}$.
W.l.o.g., assume that for every $\tau_j$, the dimension of the
smallest affine subspace of $F^d$ that contains
$\tau_j$ is exactly $q-1$
(otherwise, we add arbitrary points to  $\tau_j$).

Let $L$ be the set of all lines in $F^d$,
as in Section~\ref{sec: LDT}.
For any $\ell \in L$ and any $\tau_j$,
denote by $S(\ell,\tau_j)$ the smallest affine subspace of $F^d$
that contains both $\ell$ and~$\tau_j$.
This subspace will usually be of dimension $q+1$, and will
always be of dimension at most $q+1$.

For an assignment $a_1,...,a_{n}$ to $Y_1,...,Y_n$,
let $\tilde{A} : F^d \rightarrow F$ be the low degree extension
of $a_1,...,a_{n}$, as defined in Subsection~\ref{subsec:qlde}.
Recall that the total degree of $\tilde{A}$ is less than
$\hat{n}^{1.5}$.
For any affine subspace $S \subset F^d$, denote by $\tilde{A}|_S$
the restriction of $\tilde{A}$ to $S$. We think of
$\tilde{A}|_S$ also as a function from $F^{d'}$ to $F$, where
$d'$ is the dimension of $S$, and where formally we assume that some
(linear) parameterization of the affine space $S$ is implicit.

\subsubsection{The Correct Proof} \label{subsubsec:correct}

An assignment $a_1,...,a_{n}$ to $Y_1,...,Y_n$,
that satisfies all predicates, translates
into $(|\Phi \rangle, p)$ that causes the verifier to accept
with probability 1.
We refer to that $(|\Phi \rangle, p)$ as the {\em correct proof}.

The state $|\Phi \rangle$ of the correct proof will be
the quantum low degree extension of
$a_1,...,a_{n}$, i.e.,
$|\Psi(a_1,...,a_{n}) \rangle $, as defined in
Subsection~\ref{subsec:qlde}. Note that this is a state
of $O(\hat{n}) = O(\log n)$ qubits.
The array $p$  will have one block
for every pair $(\tau_j,S)$, such that $S \subset F^d$ is a $q+1$
dimensional affine subspace that contains $\tau_j$.
In the correct proof, this block will contain the
restriction $\tilde{A}|_{S}$
(as in~\cite{DFKRS}).
Note that the number of blocks is bounded by $k \cdot |F|^{2d}$
which is polynomial in $n$, and the size of each block
(i.e., the number of bits it takes to describe each
$\tilde{A}|_{S}$) is bounded by
$a \cdot (\hat{n}^{1.5})^{q+1}$, which is
poly-logarithmic in~$n$.

\subsubsection{The Verification Procedure} \label{subsubsec:ver}

Denote by $p(\tau_j,S)$ the content of the block indexed
by $(\tau_j,S)$ of $p$. Recall that in the correct proof
$p(\tau_j,S) = \tilde{A}|_{S}$ and recall that
$\tilde{A}|_{S}$ is a polynomial of total degree
at most $\hat{n}^{1.5}$.
Note that from $\tilde{A}|_{S}$ one can induce the
restriction of $\tilde{A}$ to $\tau_j$, as well as
the restriction of $\tilde{A}$ to any
line $\ell$ contained in $S$.
Note that the restriction of $\tilde{A}$ to $\tau_j$
gives the values of the assignment $a_1,...,a_{n}$ to
all the variables in~$t_j$, by
$a_{i} = \tilde{A}(\pi^{-1}(i))$ (for every $i \in t_j$).
Given these values, one can check wether or not the
predicate $\varphi_j$ is satisfied.
Note that in the correct proof $\varphi_j$ must be satisfied.

In general, the verifier expects $p(\tau_j,S)$ to be
a function from $S$ to $F$ (and we can think of this
function also as a function from $F^{q+1}$ to $F$).
Whenever the verifier reads a block
$p(\tau_j,S)$, the verifier can check that $p(\tau_j,S)$
is indeed a polynomial of total degree
at most $\hat{n}^{1.5}$,
and that the values induced from $p(\tau_j,S)$
to the set $\tau_j$ satisfy the predicate $\varphi_j$.
Since the verifier rejects automatically whenever
a block that doesn't pass these tests is read,
we can assume w.l.o.g.
that all the blocks pass these tests.
That is, every $p(\tau_j,S)$ is
a polynomial of total degree $\hat{n}^{1.5}$,
and the values induced from it
to the set $\tau_j$ satisfy the predicate $\varphi_j$.

The verification procedure goes as follows.
The verifier performs Step I of the quantum low degree test,
as described in Subsection~\ref{subsec: qldt}, and proceeds
to Step II.
At the beginning of Step II, the verifier needs to read
$g_{\ell}$ (for the line $\ell$ obtained in Step I).
This is done as follows:
The verifier chooses a random $j \in [k]$.
If $S(\ell,\tau_j)$ is of dimension $q+1$, define
$S=S(\ell,\tau_j)$. Otherwise, define $S$ to be a random
$q+1$ dimensional affine
subspace that contains $S(\ell,\tau_j)$.
The verifier reads $p(\tau_j,S)$
(and performs the above mentioned tests on $p(\tau_j,S)$),
and define $g_{\ell}$ to be the polynomial induced
from $p(\tau_j,S)$
(i.e., the restriction of $p(\tau_j,S)$ to $\ell$).
The verifier continues with Step II of the quantum
low degree test,
as described in Subsection~\ref{subsec: qldt}.

\subsubsection{Complexity of the Verifier}

We only need the complexity of the verifier to be
polynomial in $n$. Note, however, that if the verifier
has random access to $x,p$,
then all tasks in the
verification procedure can be performed in time
poly-logarithmic in $n$.
While this is not important for the proof, it is essential
for scaling up the proof to $NEXP$, and may be important
for possible
future applications.

\subsubsection{Analysis of the Test} \label{sussubsec: qpcp annalysis}

Denote by $V(x,|\Phi \rangle,p)$ the output of the verifier
on inputs $(x,|\Phi \rangle,p)$. Note that
$V(x,|\Phi \rangle,p)$ is a random variable.

The completeness of the test is straightforward.
If there exists an assignment $a_1,...,a_{n}$ to
$Y_1,...,Y_n$ that satisfies all predicates
$\varphi_1,...,\varphi_k$
then the correct proof $(|\Phi \rangle,p)$, as described in
Subsection~\ref{subsubsec:correct},
satisfies
$$\P[V(x,|\Phi \rangle,p) = accept] =1.$$

The soundness of the test
is given by the following
lemma.
The lemma shows that if for some
$(|\Phi \rangle,p)$, $\P[V(x,|\Phi \rangle,p) = accept]$
is large
then there is an assignment to $Y_1,...,Y_n$ that satisfies
many of the predicates $\varphi_1,...,\varphi_k$.

\begin{lemma} \label{lemma:qpcp-soundness}
Assume that for some
$(|\Phi \rangle,p)$ and some $\gamma$,
$$\P[V(x,|\Phi \rangle,p) = accept] \geq \gamma >
\frac{c' \cdot \hat{n}^{1.5}} {|F|^{\epsilon}}$$
where $c'$ is a (large enough) universal constant and
$\epsilon > 0$ is a (small enough) universal constant
(as in Lemma~\ref{lemma:qldt-sound+}).
Then,
there exists an assignment $a_1,...,a_{n}$ to
$Y_1,...,Y_n$ that satisfies at least  $\gamma^4/100$
fraction of the  predicates $\varphi_1,...,\varphi_k$.
\end{lemma}

\subsubsection{Proof of Lemma~\ref{lemma:qpcp-soundness}}

Recall that w.l.o.g. we can assume that every
$p(\tau_j,S) : S \rightarrow F$ is
a polynomial of total degree at most~$\hat{n}^{1.5}$,
and that the values induced from it
to the set $\tau_j$ satisfy the predicate $\varphi_j$.
For every line
$\ell \in L$ that is contained
in $S$,
denote  by $p(\tau_j,S)|_{\ell} : \ell \rightarrow F$
the restriction of $p(\tau_j,S)$ to $\ell$.

For every $\ell \in L$, and every $j \in [k]$,
define $g_{\ell,j}: \ell \rightarrow F$ as follows:
If $S(\ell,\tau_j)$ is of dimension $q+1$, define
$S=S(\ell,\tau_j)$. Otherwise, define $S$ to be a random
$q+1$ dimensional affine
subspace that contains $S(\ell,\tau_j)$.
Define $g_{\ell,j}$ to be
the restriction of $p(\tau_j,S)$ to $\ell$, that is,
$g_{\ell,j} = p(\tau_j,S)|_{\ell}$.
Note that $g_{\ell,j}$ is the same as the
polynomial $g_{\ell}$ defined
by the verification procedure
(see Subsection~\ref{subsubsec:ver}) in case that $j$
is the random index in $[k]$ that was picked by the procedure.
For every $\ell \in L$,
define $g_{\ell}: \ell \rightarrow F$ as follows:
Choose a random $j \in [k]$ and fix
$g_{\ell} = g_{\ell,j}$.
Note that $g_{\ell}$ is the same as the
polynomial $g_{\ell}$ defined
by the verification procedure
(see Subsection~\ref{subsubsec:ver}).

Denote, $G = \{g_{\ell}\}_{\ell \in L}$. For every
$j \in [k]$, denote
$G_j = \{g_{\ell,j}\}_{\ell \in L}$.
Recall the definition of $\Agr[h,G]$ in Subsection~\ref{subsec:cldt}.
By Lemma~\ref{lemma:qldt-sound+}
(under the assumption that $c'$ is large enough and $\epsilon$
is small enough),
there exists $h: F^d \rightarrow F$ of
total degree $\hat{n}^{1.5}$, such that,
$$\Agr[h,G] \geq \gamma^{4}/ 50.$$
By the definitions of $G,G_j$,
$$\E_j \Agr[h,G_j] = \Agr[h,G].$$
Hence, for at least $\gamma^{4}/ 100$ fraction
of the indices $j \in [k]$,
\begin{eqnarray} \label{eqn:agg}
\Agr[h,G_j] \geq \gamma^{4}/ 100.
\end{eqnarray}

Define the assignment $a_1,...,a_{n}$ to
$Y_1,...,Y_n$ to be the assignment induced from $h$, that is,
$\forall i$,
$a_{i} = h(\pi^{-1}(i))$.
We will show that for every $j \in [k]$ that satisfies
inequality~\ref{eqn:agg}, the assignment $a_1,...,a_{n}$
satisfies the predicate $\varphi_j$.
Hence, the assignment $a_1,...,a_{n}$ to
$Y_1,...,Y_n$ satisfies
at least  $\gamma^4/100$
fraction of the  predicates $\varphi_1,...,\varphi_k$,
and the lemma is proved.

\begin{claim} \label{claim:any}
For every $j \in [k]$ that satisfies
inequality~\ref{eqn:agg}, the assignment $a_1,...,a_{n}$
to $Y_1,...,Y_n$
satisfies the predicate $\varphi_j$.
\end{claim}

\proof
Fix $j \in [k]$ that satisfies
inequality~\ref{eqn:agg}.
Denote by $L' \subset L$ the set of all lines $\ell$, such that
$S(\ell,\tau_j)$ is of dimension exactly $q+1$.
Recall that for every $\ell \in L'$,
$$g_{\ell,j} = p(\tau_j,S(\ell,\tau_j))|_{\ell}$$

Since the dimension of the
smallest affine subspace of $F^d$ that contains
$\tau_j$ is $q-1 < d-2$, most lines in $L$ are also in $L'$.
More precisely,
the ratio $|L'|/|L|$ is larger than
$1-|F|^{-1}$.
Hence, by inequality~\ref{eqn:agg},
\begin{eqnarray*}
\E_{\ell \in L'}\Agr[h,g_{\ell,j}] >
\E_{\ell \in L}\Agr[h,g_{\ell,j}] - |F|^{-1}
\end{eqnarray*}
\begin{eqnarray} \label{eqn:agg2}
=
\Agr[h,G_j] - |F|^{-1} \geq \gamma^{4}/ 100 - |F|^{-1}.
\end{eqnarray}

Denote by ${\cal S}$ the set of all $q+1$ dimensional
affine subspaces
$S \subset F^d$ that contain $\tau_j$.
For every $S \in {\cal S}$, denote by
$L_S$ the set of all lines
$\ell \in L$ that are contained
in $S$, and
denote by $L'_S \subset L_S$ the set of all lines $\ell \in L$,
such that
$S(\ell,\tau_j)= S$.
In other words, $L'_S =  L_S \cap L'$.
Note that $\{ L'_S \}_{S \in {\cal S}}$ is
a partition of $L'$.
Hence, by inequality~\ref{eqn:agg2},
\begin{eqnarray*}
\E_{S \in {\cal S}} \E_{\ell \in L'_S} \Agr[h,g_{\ell,j}] =
\E_{\ell \in L'}\Agr[h,g_{\ell,j}]
\end{eqnarray*}
\begin{eqnarray} \label{eqn:agg3}
\geq \gamma^{4}/ 100 - |F|^{-1}.
\end{eqnarray}
Note that for every $\ell \in L'_S$, we have
$g_{\ell,j} = p(\tau_j,S(\ell,\tau_j))|_{\ell} = p(\tau_j,S)|_{\ell}$.
Hence,
\begin{eqnarray*}
\E_{S \in {\cal S}} \E_{\ell \in L'_S} \Agr[h,g_{\ell,j}] =
\end{eqnarray*}
\begin{eqnarray} \label{eqnequiv}
\E_{S \in {\cal S}} \E_{\ell \in L'_S}
\P_{z \in \ell} [h(z) = p(\tau_j,S)|_{\ell}(z) ].
\end{eqnarray}
For every $S \in {\cal S}$,
the dimension of the
smallest affine subspace of $S$ that contains
$\tau_j$ is $q-1 = (q+1)-2$. Hence,
most lines in $L_S$ are also in $L'_S$.
More precisely,
the ratio $|L'_S|/|L_S|$ is larger than
$1-2|F|^{-1}$.
Therefore, for every $S \in {\cal S}$,
\begin{eqnarray*}
\E_{\ell \in L_S}
\P_{z \in \ell} [h(z) = p(\tau_j,S)|_{\ell}(z)]
\geq
\end{eqnarray*}
\begin{eqnarray*}
\E_{\ell \in L'_S}
\P_{z \in \ell} [h(z) = p(\tau_j,S)|_{\ell}(z)]
- 2|F|^{-1}
\end{eqnarray*}
Hence, by inequality~\ref{eqn:agg3} and equality~\ref{eqnequiv},
\begin{eqnarray*}
\E_{S \in {\cal S}} \E_{\ell \in L_S}
\P_{z \in \ell} [h(z) = p(\tau_j,S)|_{\ell}(z)]
\geq
\end{eqnarray*}
\begin{eqnarray*}
\E_{S \in {\cal S}} \E_{\ell \in L'_S}
\P_{z \in \ell} [h(z) = p(\tau_j,S)|_{\ell}(z)]
- 2|F|^{-1}
\end{eqnarray*}
\begin{eqnarray*}
=
\E_{S \in {\cal S}} \E_{\ell \in L'_S} \Agr[h,g_{\ell,j}]
- 2|F|^{-1}
\end{eqnarray*}
\begin{eqnarray*}
\geq \gamma^{4}/ 100 - 3|F|^{-1}.
\end{eqnarray*}
Hence, by a uniformity argument,
\begin{eqnarray*}
\E_{S \in {\cal S}} \P_{z \in S} [h(z) = p(\tau_j,S)(z) ] =
\end{eqnarray*}
\begin{eqnarray*}
\E_{S \in {\cal S}} \E_{\ell \in L_S}
\P_{z \in \ell} [h(z) = p(\tau_j,S)|_{\ell}(z)]
\end{eqnarray*}
\begin{eqnarray*}
\geq \gamma^{4}/ 100 - 3|F|^{-1}.
\end{eqnarray*}
Hence, there exists (at least one) $S \in {\cal S}$, such that,
\begin{eqnarray*}
\P_{z \in S} [h(z) = p(\tau_j,S)(z)]
\geq \gamma^{4}/ 100 - 3|F|^{-1}.
\end{eqnarray*}

Recall that $h: F^d \rightarrow F$  and
$p(\tau_j,S) : S \rightarrow F$ are both polynomials of total degree
at most~$\hat{n}^{1.5}$. Thus, by Schwartz-Zippel's lemma,
if they agree on a fraction larger than
$\hat{n}^{1.5} / |F|$ of the points $z \in S$ they must agree
on every point  $z \in S$.
Thus, under the assumption that the constant $c$
(that determines the size of the field $F$) is large enough,
$h$ and $p(\tau_j,S)$ agree on every point $z \in S$.
(Note that we have the freedom to fix $c$ as large as we want).

Since we assumed that the
values induced from $p(\tau_j,S)$
to the set $\tau_j$ satisfy the predicate~$\varphi_j$,
we conclude that the
values induced from $h$
to the set $\tau_j$ satisfy the predicate $\varphi_j$.
That is, the assignment $a_1,...,a_{n}$
to $Y_1,...,Y_n$
satisfies the predicate $\varphi_j$.

This ends the proof of Claim~\ref{claim:any}.
\QED

Since inequality~\ref{eqn:agg} holds for at least
$\gamma^{4}/ 100$ fraction
of the indices $j \in [k]$,
the assignment $a_1,...,a_{n}$ to
$Y_1,...,Y_n$ satisfies
at least  $\gamma^4/100$
fraction of the  predicates $\varphi_1,...,\varphi_k$.

This ends the proof of Lemma~\ref{lemma:qpcp-soundness}.
\QED

\subsubsection{Completing the Proof of Theorem~\ref{theorem:qpcp}}
We have constructed an $(O(log n),polylog(n))$-verifier $V$,
such that on an instance
$x=(\varphi_1,...,\varphi_k)$
of $GAP(s,q,\epsilon)$ the following properties are satisfied:
\begin{enumerate}
\item
If there exists an assignment to
$Y_1,...,Y_m$ that satisfies all predicates
$\varphi_1,...,\varphi_k$,
then there
exist $|\Phi \rangle$ and $p$, such that
$$\P[V(x,|\Phi \rangle,p) = accept] =1.$$
(See Subsection~\ref{sussubsec: qpcp annalysis}).
\item
If any assignment to $Y_1,...,Y_m$ satisfies
at most $\epsilon$ fraction of the predicates
$\varphi_1,...,\varphi_k$, then for
any $|\Phi \rangle$ and $p$,
$$\P[V(x,|\Phi \rangle,p) = accept] \leq o(1).$$
(By Lemma~\ref{lemma:qpcp-soundness}).
\end{enumerate}
Hence, $GAP(s,q,\epsilon) \in \QPCP[log (n),polylog(n),o(1)]$,
and since $GAP(s,q,\epsilon)$ is $NP$-complete we conclude that
$NP \subset \QPCP[log (n),polylog(n),o(1)]$.
\QED

\subsection*{Acknowledgment}

I am grateful to Adam Smith for simplifying the retrieval
protocol of Subsection~\ref{subsec:retrieval}
(and for allowing me to include here the simplified version),
and to Amir Shpilka and Yael Tauman Kalai
for very helpful conversations.

\end{document}